\documentclass[reprint,amsmath,amssymb,aps]{revtex4-1}
\usepackage[utf8]{inputenc}
\usepackage[english]{babel}
\usepackage[babel]{csquotes}
\usepackage{stmaryrd}
\usepackage{amssymb}
\usepackage{amsmath}
\usepackage{graphicx}
\providecommand{\eps}{\epsilon}

\DeclareMathOperator{\const}{const}

\begin{document}

%\begin{frontmatter}
\title{Cosmological Inflation from the dark Universe}

\author{Christian Henke} 
\email{henke@math.tu-clausthal.de}
%\institute{University of Technology at Clausthal, \email{henke@math.tu-clausthal.de},\\ Department of Mathematics,\\ Erzstrasse 1, D-38678 Clausthal-Zellerfeld, Germany}
\affiliation{Department of Mathematics, University of Technology at Clausthal, \\ Erzstrasse 1, D-38678 Clausthal-Zellerfeld, Germany}

\date{\today}

\begin{abstract}
The present work deals with the cosmological consequences of the variable cosmological term $\Lambda(a)=\Lambda_0 + \Lambda_1 a^{-r} + \Lambda_2 a^{-s},$ where $a$ is the scale factor of the Robertson-Walker space-time.
An analysis of the parameters $\Lambda_1,\Lambda_2,r,s$ and $\alpha,$ which describes the interaction between the cosmological constant and non-relativistic matter/radiation, leads to an attractive density $\rho_{\text{dm}} \sim a^{-r}$ and a repulsive density $\rho_{\text{infl}} \sim a^{-s}.$
%The covariant conservation of the energy-momentum tensor, the intrinsic coupling parameter $\alpha$ which describes the interaction between the cosmological constant and non-relativistic matter/radiation and the parameter ranges $3 <r<12/(4-\alpha),\Lambda_1<0$ and $4<s, 0<\Lambda_2$ leads to an attractive density $\rho_{\text{dm}} \sim a^{-r}$ and a repulsive density $\rho_{\text{infl}} \sim a^{-s}.$ 
As already known from literature, an approach of the form $\Lambda(a)=\Lambda_0 + \Lambda_1 a^{-r}$ and the identification of the energy density of space-time expansion with the zero-point density can avoid the scale discrepancy of the cosmological constant problem and solve the missing mass problem of dark matter. 
%The identification of the energy density of space-time expansion in an empty universe with the zero-point density and a more accurate analysis of the parameter $\Lambda_1,r$ and $\alpha$ demonstrates that the scale discrepancy of the cosmological constant problem is avoided and that the missing mass problem of dark matter is solved.
Furthermore, this works demonstrates that the additional term $\Lambda_2 a^{-s}$ solves the flatness, the horizon and the magnetic-monopole problem without a hypothetical scalar field, which avoids speculations about the origin of this field. 
As a consequence, the imaginary time concept from 
Hartle and Hawking (cf. \cite{Hartle.Hawking_Wavefunction_1983})
appears naturally.
%As the consequence, the model under consideration is compatible with observations of rotational curves of galaxies and with constraints from the standard model of particle physics.
%Moreover, there exists a time $t_1$ such that the time before that becomes imaginary. The adaption of $\Lambda_2$ and $s$ such that $t_1$ equals the Planck time solves the flatness, the horizon and the magnetic-monopole problem without a hypothetical scalar field, which avoids speculations about the origin of this field. 
%These macroscopic conclusions agree with the quantum arguments introduced by Hartle and Hawking (cf. \cite{Hartle.Hawking_Wavefunction_1983}).
\end{abstract}

\maketitle
%\end{frontmatter}

\section{Introduction}
In the recent paper \cite{Henke_TheCosmologicalNature_2018}, the author has proved that Friedmann equations and the covariant conservation of the ener\-gy-momentum tensor which couples only a small contribution of $\Lambda(a)=\Lambda_0+\Lambda_1 a^{-r}, 3 <r<3.0000075$ with non-relativistic matter avoids the scale discrepancy of the cosmological constant problem 
(see
\cite{Carroll_TheCosmologicalConstant_1992,Carroll_TheCosmologicalConstant_2001,Sahni.Starobinsky_TheCasefora_2000,Weinberg_TheCosmologicalConstant_1989,Dolgov_ProblemVacuumEnergy_1997}).
Moreover, it has been demonstrated that this approach extends the $\Lambda$CDM model (cf. \cite{Ade_Planck_2015})
such that the missing mass problem of dark matter is solved, observations of rotational curves of galaxies are satisfied 
 and constraints from the standard model of particle physics are taken into account (cf. \cite{Barranco.Bernal.ea_DarkMatterEquationState_2015},\cite{Freese.Adams.ea_CosmologyDecayingVacuumEnergy_1987}).
Here, $\Lambda_0$ represents the dark energy and $\Lambda_1 a^{-r}$ is the generator of cold and hot dark matter. 

The novelty of this work is to present an inflation theory which is caused by a cosmological term $\Lambda_2 a^{-s}$ and avoids the use of a hypothetical scalar field with arbitrary potentials. Furthermore, this expands the theory of imaginary time from Hartle and Hawking with a macroscopic interpretation.

%As well as the covariant conservation of the energy-momentum tensor creates and annihilates dark matter from $\Lambda_1 a^{-r},$ dark inflation should be created and annihilated from $\Lambda_2 a^{-s}.$

The remainder of the paper is organised as follows. In section 2 we review the Friedmann equations where the cosmological term $\Lambda$ is a function of the scale factor $a$. 
The variable cosmological term and the identification of the energy density of space-time expansion with the zero-point density which is also considered in section 3 has proven to be a very fruitful starting point for the explanation of the scale discrepancy causing the cosmological constant problem (cf.  \cite{Henke_TheCosmologicalNature_2018}, \cite{Henke_QuantumVacuumEnergy_2018}).
%Section 3 is devoted to the dark sector and states the basic ideas behind the derivation of the cancellation mechanism of the cosmological constant problem from .
Then, the gravitational nature of $\rho_{\text{dm}} \sim a^{-r}$ and $\rho_{\text{infl}} \sim a^{-s}$ are investigated. Section 4 derives the concept of imaginary time from Friedmann's equation and the properties of $\rho_{\text{infl}}$ and shows how the key problems of the ordinary Big Bang singularity are solved.
Finally, we show that our solution of the cosmological constant problem satisfies the constraints of the imaginary time concept and that the derived creation and annihilation process of dark matter agrees with the standard model of cosmology.

\section{A time-dependent cosmological term}

Let us consider a homogeneous and isotropic universe by using the Robertson-Walker space-time line element 
(see \cite{Wald_GeneralRelativity_1984} for notational conventions)
\begin{equation*}
ds^2=-c^2 dt^2+a(t)^2 \left(\frac{dr^2}{1-kr^2} +r^2 \left( d\theta^2 + \sin^2 \theta d \phi^2\right)\right),
\end{equation*}
which reduces Einstein's field equations to Friedmann's equations
for the scaling factor $a(t)$
\begin{align}
\frac{\dot{a}^2}{a^2} -  \frac{1}{3} \Lambda +\frac{k}{a^2}&= \frac{\kappa c^2}{3}  \rho,
\label{eq:friedmann0}\\
3\frac{ \ddot{a}}{a} - \Lambda  &=-\frac{\kappa}{2} \left( \rho c^2 + 3p \right).
\label{eq:friedmann1}
\end{align}
For the derivation of $(\ref{eq:friedmann0})$ and $(\ref{eq:friedmann1})$, it makes no difference whether one takes $\Lambda=\const$ or $\Lambda=\Lambda(t).$

In order to fulfill the Bianchi identities and assume that the conservation of energy-momentum tensor of matter holds, one can propose a time-dependent gravitational term $G$ and link the derivatives of $G$ and $\Lambda$ (cf.  \cite{Abdussattar.Vishwakarma_SomeFRWmodels_1997}). 
In this paper, a different approach is considered which leaves the form of Einstein's equation formally unchanged and also satisfies the Bianchi identities (cf. \cite{Overduin_Evolutionofthe_1998}). 
That is, the $\Lambda$-term is included in the energy-mo\-men\-tum tensor of the right-hand side
\begin{equation*}
T^{\text{eff}}_{\mu \nu}=\left(\rho_\text{eff} + \frac{p_\text{eff}}{c^2}\right) u_\mu u_\nu + p_\text{eff} \,g_{\mu \nu},
\end{equation*}
where we define an effective density and pressure field which also includes non-relativistic matter and radiation
\begin{equation*}
\rho_{\text{eff}}=\rho_m + \rho_r + \rho_\Lambda, \quad p_{\text{eff}}=p_m + p_r +p_\Lambda.
\label{eq:eff_rho_p}
\end{equation*}
Here, $\rho_\Lambda=\Lambda(t)/\kappa c^2$ and $p_\Lambda=-\Lambda(t)/\kappa$ denote the cosmological fields. 
By the transformation properties of $T^\text{eff}_{\mu \nu}$, Einstein's field equation with $\Lambda(t)$ is still independent of the coordinates and therefore perfectly consistent with covariance (cf. \cite{Shapiro.Sola_Possiblerunningcc_2009}). 
Moreover, the right-hand side interpretation of $\Lambda(t)$, ensures the existence of an action that reproduces Einstein's field equation with a running cosmological constant.  
According to \cite[p. 2]{Overduin_Evolutionofthe_1998}, there are no a priori reasons why $\Lambda$ should not vary - as long as the covariant energy conservation 
\begin{equation}
\nabla^\mu T^\text{eff}_{\mu \nu}=0,
\label{eq:cov_cons}
\end{equation}
of the effective tensor is satisfied, which in turn implies the Bianchi identities. 

As mentioned in the introduction, the variable cosmological term is defined by 
\begin{equation*}
\Lambda(t)=\Lambda(a(t))=\Lambda_0 + \Lambda_1 a^{-r}(t)+\Lambda_2 a^{-s}(t),\quad \Lambda_0,r,s,>0.
\label{eq:lambda_model}
\end{equation*}
An approach of the form $\sim{a^{-r}}$ belongs to models whose theoretical origin is currently unknown. Nevertheless, there are numerous reasons why such scenarios are considered in the literature (see, for examples, \cite{Henke_QuantumVacuumEnergy_2018}, \cite{Henke_TheCosmologicalNature_2018}, \cite{Overduin_Evolutionofthe_1998} and the references therein). 

Moreover, the covariant conservation of the energy-momen\-tum tensor $(\ref{eq:cov_cons})$ gives
\begin{equation*}
\frac{d}{da} \left(\rho_{\text{eff}} a^3 \right) +3 \frac{p_{\text{eff}}}{c^2} a^2=0, \quad p'_\text{eff}=0,
\label{eq:divT_rho_p}
\end{equation*}
and leads with $p_m=0$ and $p_r=c^2 \rho_r/3$ to
\begin{equation*}
\frac{1}{a^3}\frac{d}{da} \left(\rho_m a^3 \right) 
+\frac{1}{a^4}\frac{d}{da} \left(\rho_r a^4 \right) 
=- \frac{1}{\kappa c^2} \frac{d}{da} \Lambda(a),
\label{eq:divT_rho_p2}
\end{equation*}
which is satisfied by the parametrised coupling
\begin{align*}
\frac{1}{a^3}\frac{d}{da} \left(\rho_m a^3 \right) 
&=- \alpha \frac{1}{\kappa c^2} \frac{d}{da} \Lambda(a),\\
\frac{1}{a^4}\frac{d}{da} \left(\rho_r a^4 \right) 
&=- (1-\alpha) \frac{1}{\kappa c^2} \frac{d}{da} \Lambda(a),\quad 0 <\alpha <1.
\end{align*}
Integrating both equations gives for $r,s \neq 3$
\begin{equation}
\rho_m= \rho_{m,0} a^{-3}
+ \frac{\alpha r \Lambda_1}{\kappa c^2}\frac{1}{3-r} a^{-r}
+ \frac{\alpha s \Lambda_2}{\kappa c^2}\frac{1}{3-s} a^{-s},\label{eq:rho_m}
\end{equation}
and
\begin{equation}
\rho_r= \rho_{r,0} a^{-4}
+\frac{(1-\alpha) r \Lambda_1}{\kappa c^2}\frac{1}{4-r} a^{-r}
+\frac{(1-\alpha) s \Lambda_2}{\kappa c^2}\frac{1}{4-s} a^{-s},
\label{eq:rho_r}
\end{equation}
for $r,s \neq 4.$
Hence, the interaction between the variable cosmological term with non-relativistic matter and radiation generates additional matter components.

\section{The dark sector}
In this section we review the basic ideas of the cancellation mechanism between the quantum zero-point energy and the energy density of the cosmological constant in an empty universe (cf. \cite{Henke_QuantumVacuumEnergy_2018}). First, using a metric which is independent of the scale factor, the second Friedmann equation can be identified with an Euler-Lagrange equation. Therefore, the related Lagrangian and the first Friedmann equation leads to the cancellation mechanism 
between the energy density of space-time expansion and the energy density of the cosmological term. 
Identifying the energy density of space-time expansion with the quantum zero-point energy, % $(\rho_\text{zpe}=6 \rho_a/a^4),$
we get for $r,s\neq 4$ 
\begin{equation}
\rho_{\Lambda}= \rho_{\text{zpe}}+\frac{4 \int^a \Lambda(\alpha) \alpha^3 \, d\alpha}{\kappa c^2 a^4},
\label{eq:cancellation_mech}
\end{equation}
where
\begin{equation*}
\rho_{\text{zpe}}
%&=
%\frac{1}{\kappa c^2} \left( \Lambda(a)-\frac{4}{a^4} \int \Lambda(a) a^3 \, da \right) \\
=\frac{1}{\kappa c^2} \left( \Lambda_1\frac{r}{r-4} a^{-r}
+ \Lambda_2\frac{s}{s-4} a^{-s} \right) \overset{!}{=} \frac{1}{2 \pi \kappa l_p^2 c^2},
\end{equation*}
and $l_p$ denotes the Planck length.
It follows that
\begin{equation}
\Lambda_1=\frac{r-4}{r} \left( \frac{1}{2 \pi l_p^2} -\Lambda_2 \frac{s}{s-4}\right).
\label{eq:lam1}
\end{equation}
As the consequence, there is no scale discrepancy between the total energy densities and the fine-tuning problem of the cosmological constant problem is solved.

In order to analyse the gravitational nature of the matter components with the exponents $r$ and $s,$ we rearrange the density components to 
\begin{align*}
\rho_\text{eff}&= \rho_{m,0} a^{-3}+ \rho_{r,0} a^{-4}
+\frac{\Lambda_0}{\kappa c^2} 
+ \rho_{\text{dm}} + \rho_{\text{infl}},\\
\rho_{\text{dm}}&=\frac{\Lambda_1}{\kappa c^2}\left( 1+\frac{\alpha r}{3-r}+\frac{(1-\alpha)r}{4-r} \right) a^{-r}, \quad 3 \neq r \neq 4, \\
\rho_{\text{infl}}&=\frac{\Lambda_2}{\kappa c^2}\left( 1+\frac{\alpha s}{3-s}+\frac{(1-\alpha)s}{4-s} \right) a^{-s}, \quad 3 \neq s \neq 4,
\end{align*}
where the names $\rho_{\text{dm}}$ and $\rho_{\text{infl}}$ already anticipate the character of the densities.

In analogy to \cite{Henke_TheCosmologicalNature_2018}, we get 
\begin{align*}
3 \frac{\ddot{a}}{a}
&=-\frac{\kappa c^2}{2}\left( \rho_{m,0} a^{-3}  + 2 \rho_{r,0} a^{-4} \right)+ \Lambda_0\\
&-\frac{\kappa c^2}{2}(r-2) \rho_{\text{dm}} -\frac{\kappa c^2}{2}(s-2) \rho_{\text{infl}}.
\end{align*}
To get an attractive behaviour of $\rho_{\text{dm}}$ and an repulsive nature of $\rho_{\text{infl}},$ 
%which is dominant in the early universe, 
we have to demand
\begin{equation}
\rho_{\text{dm}}>0,\quad \rho_{\text{infl}}<0.
%\quad \rho_\text{eff} \approx \rho_{\text{infl}} \text{ for } t \to 0.
\label{eq:density_constr}
\end{equation}
By considering the identity
\begin{equation*}
\left( 1+\frac{\alpha r}{3-r}+\frac{(1-\alpha)r}{4-r}\right)
=\frac{\left(4-\alpha\right)\left(r-\frac{12}{4-\alpha}\right)}{(3-r)(r-4)},
\end{equation*}
we find that $(\ref{eq:density_constr})$ is satisfied if
%$|\Lambda_2|$ is sufficiently large,
\begin{equation}
3 < r < \frac{12}{4-\alpha},\quad \left( \frac{1}{2 \pi l_p^2} -\Lambda_2 \frac{s}{s-4}\right) >0,
\label{eq:param_range_r}
\end{equation}
 and
\begin{equation}
\Lambda_2>0,\quad 4<s 
\text{ or }
\Lambda_2<0,\quad \frac{12}{4-\alpha} \le s<4. 
\label{eq:param_range_s}
\end{equation}
Because of the positivity and the attractive nature of $\rho_{\text{dm}},$ it could be identified as dark matter. Moreover, the repulsive behaviour of $\rho_{\text{infl}}$ is responsible for the inflation of $a$ in the early universe if $\rho_{\text{infl}}$ dominates for $t \to 0,$ which is satisfied for an sufficiently large $|\Lambda_2|.$ 
As usual, the $\Lambda_0$-depending term acts repulsive too and is labeled as dark energy.

From $(\ref{eq:rho_m})$ and $(\ref{eq:rho_r})$ we can conclude that dark matter and inflation  don't interact with ordinary matter and radiation ($\rho_{m,0}$ and $\rho_{r,0}$ are independent of $\Lambda_1$ and $\Lambda_2$). 
%The interaction takes place between the three components of dark matter and inflation which can be classified by their equation of state parameter $w=-1, w=0$ and $w=1/3.$ 
The interaction takes place between the cosmological term, cold and hot dark matter and between the cosmological term, the cold and the hot inflation field. 

\section{Cosmological Inflation}

%\section{Cosmological constraints}
It remains to analyse the effects of the inflation density.
%It remains to consider some observational constraints for the present-day composition of the universe. 
Using the usual settings $H/c=\dot{a}/a,$
\begin{equation*}
\begin{aligned}
\Omega_k&=&-\frac{k c^2}{a_0^2 H_0^2}&,& 
\Omega_m&=& \frac{\rho_{m,0}}{\rho_{\text{crit}}} a_0^{-3},\\
\Omega_r&=& \frac{\rho_{r,0}}{\rho_{\text{crit}}} a_0^{-4}&,&
\Omega_{\Lambda_0}&=&\frac{\Lambda_0 c^2}{3 H_0^2},
\end{aligned}
\end{equation*}
where $a_0=a(t_0)$ and $H_0=H(t_0)$ denote the present-day scale factor and Hubble constant, the contributions from dark matter and inflation are taken into account by
\begin{equation*}
\Omega_{\text{dm}}=\frac{\rho_{\text{dm}}}{\rho_{\text{crit}}},\quad
\Omega_{\text{infl}}=\frac{\rho_{\text{infl}}}{\rho_{\text{crit}}},\quad
\rho_{\text{crit}}=3 H_0^2/\kappa c^4.
\end{equation*}
Multiply $(\ref{eq:friedmann0})$ by $c^2 a^2/H_0^2 a_0^2,$ we get 
\begin{equation}
\left( \frac{dx}{d\tau} \right)^2=\Omega_k +\frac{\Omega_m}{x} + \frac{\Omega_r}{x^2} +\Omega_{\Lambda_0} x^2 +\frac{\Omega_{\text{dm}}}{x^{r-2}}+\frac{\Omega_{\text{infl}}}{x^{s-2}},
\label{eq:evolution_universe}
\end{equation}
where $x=a/a_0$ and $\tau=H_0 t.$ 
Because of $\Omega_{\text{infl}}/x^{s-2}<0,$ which dominates in the early universe, there exists $\tau_1=H_0 t_1$ such that  $( dx/d\tau )^2<0$ for $\tau < \tau_1.$ 
Using the concept of imaginary time introduced by \cite{Hartle.Hawking_Wavefunction_1983}, i.e. changing $\tau \to i\tau$ for $\tau < \tau_1,$ the left hand side of $(\ref{eq:evolution_universe})$ is always positive. 
Therefore, the early universe can be described by an Euclidean space-time line element
\begin{equation*}
ds^2=c^2 dt^2+a(t)^2 \left(\frac{dr^2}{1-kr^2} +r^2 \left( d\theta^2 + \sin^2 \theta d \phi^2\right)\right).
\end{equation*}
Hence, the interval $[0,t_1]$ is spacelike, there are no timelike intervals $(ds<0)$ and no light cones $(ds=0)$ provided that $k \le0.$
 
Let $t_2$ denotes the end of inflation. The flatness, the horizon and the magnetic-monopole problem can be avoid\-ed if
the number of e-folds during inflation satisfies the inequality (cf. \cite{Chongchitnan_Inflationmodelbuilding_2016})
\begin{equation}
\tilde{N}=\ln \left(\frac{a(t_2) H(t_2)}{a(t_1) H(t_1)}\right)
%= \frac{\frac{d}{d\tau} x(\tau_2)}{\frac{d}{d\tau} x(\tau_1)} 
\succsim 60.
\label{eq:efolds}
\end{equation}
Since $d^2x(\tau_2)/d\tau^2 =0$ and $d^2x(\tau)/d\tau^2 >0$ for $\tau < \tau_2,$ $dx(\tau)/d\tau $ has a finite maximum at $\tau=\tau_2.$  
Moreover, $d x(\tau_1)/d\tau=0$ yields 
\begin{equation*}
\frac{a(t_2) H(t_2)}{a(t_1) H(t_1)}
= \frac{\frac{d}{d\tau} x(\tau_2)}{\frac{d}{d\tau} x(\tau_1)} =\infty,
\end{equation*}
which fulfills equation $(\ref{eq:efolds}).$

\section{Cosmological constraints}
In this section we investigate to what extent our present cosmos can be described with the presented variable cosmological term.
Equation $(\ref{eq:evolution_universe})$ leads to the present-day constraint
\begin{equation*}
1=\Omega_k + \Omega_m + \Omega_r + \Omega_{\Lambda_0}+ \Omega_{\text{dm}}+ \Omega_{\text{infl}}.
\end{equation*}
To relate the last equation with observations (cf. \cite{Ade_Planck_2015}), we consider 
$\Omega_k=0,\Omega_m=0.05,\Omega_r=5\times 10^{-5},\Omega_{\Lambda_0}+\Omega_{\text{infl}}=0.69,\Omega_{dm}=0.26$
and $H_0=67.74 \frac{km}{s\, Mpc}.$ 
The setting of $\Omega_{\text{infl}}$ will be discussed further down.
Moreover, the definition of $\Omega_{\Lambda_0}$ leads to a cancellation mechanism $(\ref{eq:cancellation_mech})$ which automatically cancel 121 decimal places without fine-tuning.
First, let us discuss the inflation constraints.
The beginning of the inflation $\tau_1$ defines the first time of the universe. Therefore, the setting $\tau_1=H_0 t_p,$ where $t_p$ denotes the Planck time, is used. As a consequence and in contrast to the usual Big Bang theories, the approach under consideration is valid for $t< t_p.$ 

The definition of $\tau_1$ reduces the degree of freedom of the inflation parameters to one, i.e. $\Omega_{\text{infl}}=\Omega_{\text{infl}}(s)$ and therefore $x_1=x_1(s).$ In order to choose the parameter $s$ one can consider $e.g.$ the scalar spectral index $n_s$ and the tensor to scalar power ratio $r$ (cf. \cite{Ade_Planck_inflation_2015}). 
Usually, $n_s$ and $r$ can be determined in the regime $0 < \epsilon_1 \ll 1, \epsilon_1=- \frac{d}{dt} H/H^2$ by the Hubble flow functions (see \cite{Ade_Planck_inflation_2015} and the references therein). As the consequence of 
\begin{equation*}
\frac{\ddot{a}}{a}=\frac{\dot{H}}{c}+\frac{H^2}{c^2}=\frac{H^2}{c^2}(1-\epsilon_1)
=\frac{\dot{a}^2}{a^2}(1-\epsilon_1),
\label{}
\end{equation*}
it follows because of
\begin{equation*}
\eps_1=1-\frac{x \frac{d^2 x}{d \tau^2} }{\left(\frac{dx}{d\tau} \right)^2},
\label{}
\end{equation*}
that the requirement is not fulfilled
\begin{equation*}
\epsilon_1(\tau_1)=-\infty,\quad \epsilon_1(\tau_2)=1.
\label{}
\end{equation*}
As an alternative, the numerical simulation of the primordial spectra of the scalar and tensor perturbations can also determine $n_s$ and $r.$ However, such a simulation is beyond the scope of the paper and we proceed with the unknown parameter $s$. 

\begin{figure}[!htbp]
\begin{center}
\includegraphics[width=1.0\columnwidth]{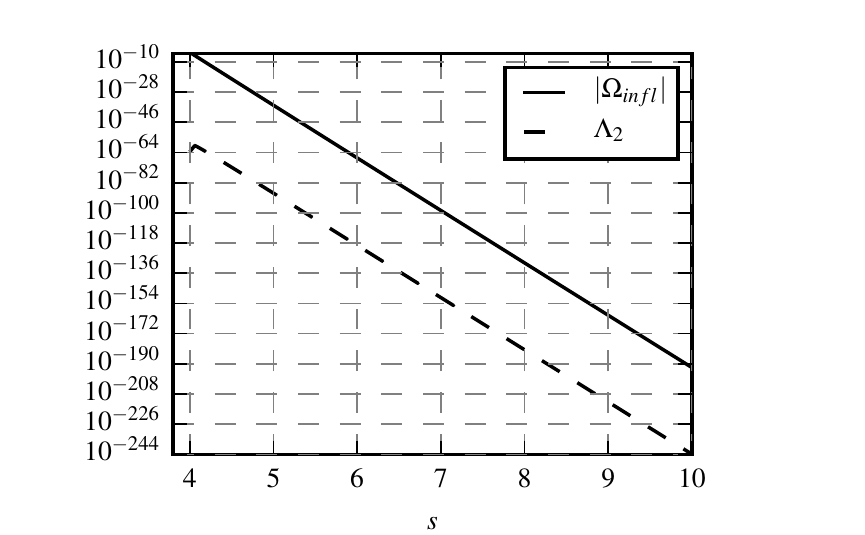}
\caption{The inflation parameters $\Omega_{\text{infl}}$ and $\Lambda_2$ with respect to $s$.}
\label{fig:OInfl_L2}
\end{center}
\end{figure}

\begin{figure}[!htbp]
\begin{center}
\includegraphics[width=1.0\columnwidth]{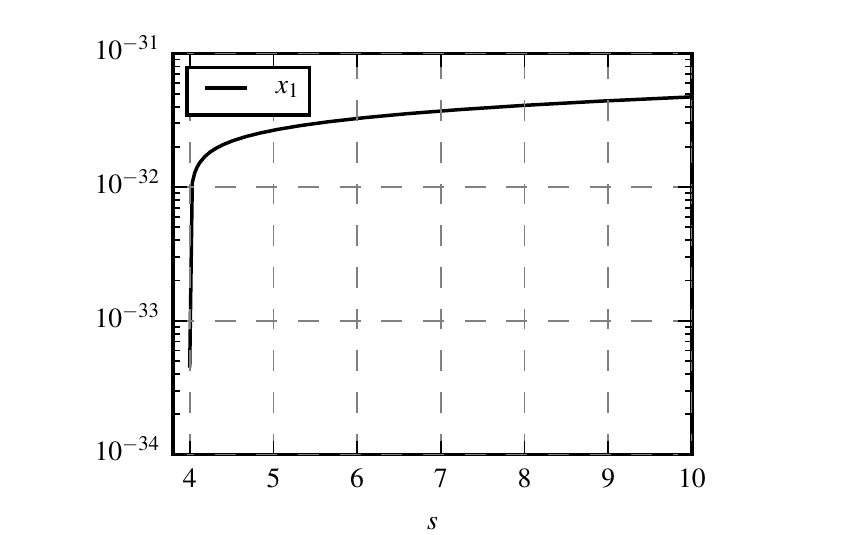}
\caption{The scaling paramter $x_1$ at the border between imaginary and real time}
\label{fig:x1}
\end{center}
\end{figure}

From the equations $(\ref{eq:density_constr})$ and $(\ref{eq:evolution_universe})$ and the definition of $x_1,$ the following small value approximations arise for the early universe
\begin{equation*}
\left( \frac{dx}{d\tau} \right)^2=\frac{\Omega_r}{x^2} -\frac{|\Omega_{\text{infl}}|}{x^{s-2}}, \quad
0=\frac{\Omega_r}{x_1^2} -\frac{|\Omega_{\text{infl}}|}{x_1^{s-2}}.
\end{equation*}
It follows that 
\begin{equation}
x_1(s)=\left( |\Omega_{\text{infl}}(s)|/\Omega_r \right)^{1/(s-4)}
\label{eq:x1_s}
\end{equation}
and
\begin{equation}
i t_p =\frac{1}{H_0} \int_0^{x_1(s)} \frac{dx}{\sqrt{\Omega_r/ x^{2}-|\Omega_\text{infl}(s)| x^{2-s}}}.
\label{eq:t1}
\end{equation}
In order to integrate equation $(\ref{eq:t1}),$ it is convenient to consider
\begin{equation*}
F(s)=\int_0^1 \frac{\lambda}{\sqrt{\lambda^{-\xi}-1}} \,d\lambda, \quad \xi=s-4.
\end{equation*}
Therefore, we can write
\begin{equation*}
\Omega_{\text{infl}}(s)=-\Omega_r \left( \frac{t_p H_0 \sqrt{\Omega_r}}{F(s)} \right)^{\frac{s-4}{2}}
\label{eq:Oinfl_s}
\end{equation*}
and
\begin{equation*}
x_1(s)=\left( \frac{t_p H_0 \sqrt{\Omega_r}}{F(s)} \right)^{\frac{1}{2}},
\label{eq:x1_s2}
\end{equation*}
(cf. Fig. \ref{fig:OInfl_L2} and Fig. \ref{fig:x1}).
Let us consider, for a moment, the case $s<4.$ Because of $iF(s) \in \mathbb{R},$ time has an imaginary character for $t>t_p.$ Therefore, the constraints in $(\ref{eq:param_range_s})$ become
\begin{equation*}
\Lambda_2>0,\quad 4<s
\label{eq:param_range_s2}
\end{equation*}
and $(\ref{eq:x1_s})$ yields 
\begin{equation}
|\Omega_\text{infl}(s)|<\Omega_r.
\label{ineq:Omega_infl}
\end{equation}
Moreover, it follows from $\xi=s-4 >0$ and with
the help of the Gaussian hypergeometric function $_2F_1(a,b,c,z)$ for $a \in \mathbb{C}, b=1/2+2/\xi, c-a=1+2/\xi, z=1$ that 
\begin{equation*}
F(s)=\frac{1}{\xi} \int_0^1 \frac{x^{\frac{4-\xi}{2 \xi}}}{\sqrt{1-x}}\, dx
=\frac{\sqrt{\pi}}{2} \frac{\Gamma(\frac{1}{2}+\frac{2}{\xi})}{\Gamma(\frac{2}{\xi})}.
\end{equation*}
For the sake of simplicity we consider the parameter range $4<s\le 10,$ since $\Omega_\text{infl}(10)=-5.628 \times 10^{-193}$ should be small enough to consider additional constraints from the primordial spectra.

Second, to consider the dark matter constraints, the term 
\begin{equation*}
\Lambda_2(s) \frac{s}{s-4}=\frac{3 H_0^2 |\Omega_\text{infl}| s (s-3)}{c^2 (4-\alpha)(s-\frac{12}{4-\alpha})}
\end{equation*} 
is discussed.
It follows from $(\ref{ineq:Omega_infl})$ that
\begin{equation*}
\Lambda_2(s) \frac{s}{s-4} \underset{4<s\le 10}{\le} \frac{30 H_0^2 \Omega_r}{4 c^2 (4-\alpha)(1-\frac{3}{4-\alpha})}=2.011 \times 10^{-56}.
\end{equation*} 
As a consequence, the second constraint of $(\ref{eq:param_range_r})$ is satisfied. Moreover, the coupling between the dark matter and inflation parameters can be neglected and equation $(\ref{eq:lam1})$ turns into  
\begin{equation*}
\Lambda_1=\frac{1}{2 \pi l_p^2} \frac{r-4}{r} .
\end{equation*}

Using
$\Omega_{\text{dm}}=0.26$ and
\begin{equation*}
\epsilon=\frac{2 \pi l_p^2 H_0^2 \Omega_{\text{dm}}}{c^2}=2.29 \times 10^{-122},
\end{equation*}
it follows in analogy to \cite{Henke_TheCosmologicalNature_2018} that
\begin{equation*}
r=3 +\frac{3 \alpha}{4-\alpha} -\frac{108 \alpha \epsilon}{(4-\alpha)^3} + \mathcal{O}(\epsilon^2),
\label{eq:r3}
\end{equation*}
which results in a $13.80$ years old universe as long as $0 < \alpha \le 10^{-5}.$Moreover, the remaining parameters of the cosmological term are $\Lambda_1=-2.03 \times 10^{68}$ and $3 <r\le 3.0000075.$ Hence, the overall cosmological term $\Lambda(a)$ is negative! 

For the consequences on constraints from the standard model of particle physics and from observations of rotational curves of galaxies see \cite{Henke_TheCosmologicalNature_2018}.

\section{Concluding remarks}
In this paper, the variable cosmological term $\Lambda(a)=\Lambda_0 + \Lambda_1 a^{-r}+\Lambda_2 a^{-s}, r,s >0$ has been applied and it has been confirmed that $\Lambda_2 a^{-s}, \Lambda_2>0,4<s$ describes the inflation of the early universe without a hypothetical scalar field which avoids speculations about its origin.
More precisely, $\Lambda_2 a^{-s}$ creates/destroys the inflation density field with the interaction of non-relativistic matter/radiation, such that the density field generates the imaginary time structure from \cite{Hartle.Hawking_Wavefunction_1983}, which happens before the universe shrinks so far as the quantum effects become dominant and the equations invalid. As a side effect, the flatness, the horizon and the magnetic-monopole problems are solved automatically. 

Moreover, it has also been confirmed that the identification of the total energy density of an empty Friedmann universe with the zero-point energy density avoids the fine-tuning problem of the cosmological constant problem. 
As a consequence, $\Lambda_1 a^{-r}$ generates the attractive force and the missing mass of dark matter.
The accepted age of our universe has explained by coupling only a small fraction of the cosmological term with non-relativistic matter. As the consequence, the parameter range shrinks to $3<r<3.0000075,$ which satisfies numerous observational constraints (see \cite{Henke_TheCosmologicalNature_2018}) and establishes a negative cosmological term in our universe.
This could have important consequences for holographic correspondence-theories which are mainly formulated on space-times with a negative cosmological constant (cf. \cite{Aharony.Gubser.ea_LargeNField_2000}) and could prevent the swampland in string theories (cf. \cite{Obied.Ooguri.ea_DeSitterSpaceSwampland_2018}). 

%\section*{References
%\bibliography{QVEGR2}
%\bibliographystyle{unsrt}
\bibliographystyle{apa}

\end{document}